\documentclass[aps,pre,preprint,floatfix]{revtex4}
\usepackage{graphicx}
\usepackage{amsmath}
\usepackage{dsfont}

\begin{document}

\title{Active Brownian Rods}


\author{Fernando Peruani}
\affiliation{Universit{\'e} Nice Sophia Antipolis, Laboratoire J.A. Dieudonn{\'e}, UMR 7351  CNRS, Parc Valrose, F-06108 Nice Cedex 02, France}

\begin{abstract}
Lecture Notes for the Summer School ``Microswimmers -- From
Single Particle Motion to Collective Behaviour''  at 
Forschungszentrum J{\"{u}}lich, 2015. 
%
%
%
%
\end{abstract}


\maketitle

\section{Introduction}

\begin{figure} [b]
\begin{center}
\resizebox{12cm}{!} {\includegraphics{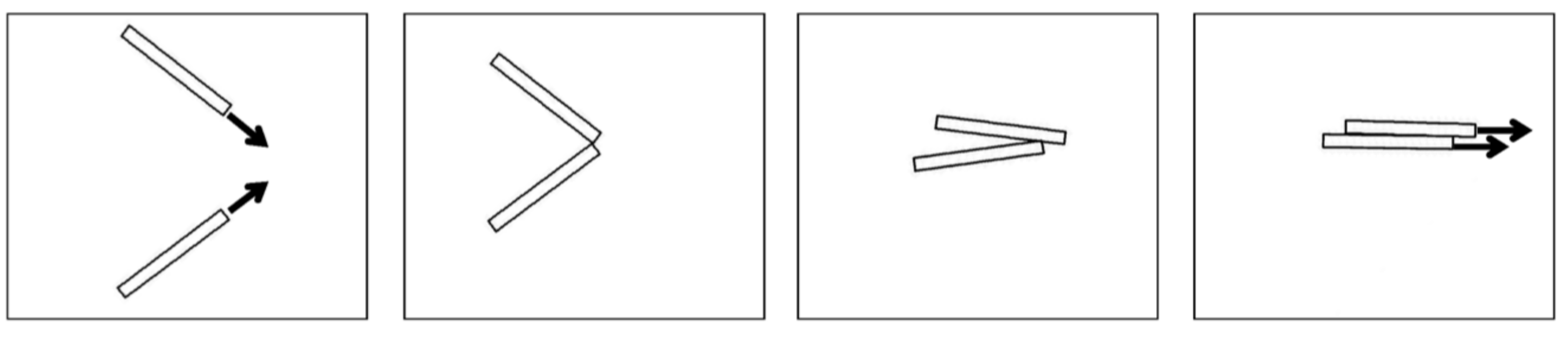}}
\caption{Sketch of two interacting rods in the physical active Brownian rod model (taken from~\cite{weitz2015}). Chronological snapshots of a collision between two rods in a simulation. Notice that even though the interaction is exclusively repulsive, it leads  to an effective velocity alignment (and an  effective attraction). 
}
\label{fig-model}
\end{center}
\end{figure}

\begin{figure} 
\begin{center}
\resizebox{10cm}{!}{\rotatebox{0}{\includegraphics{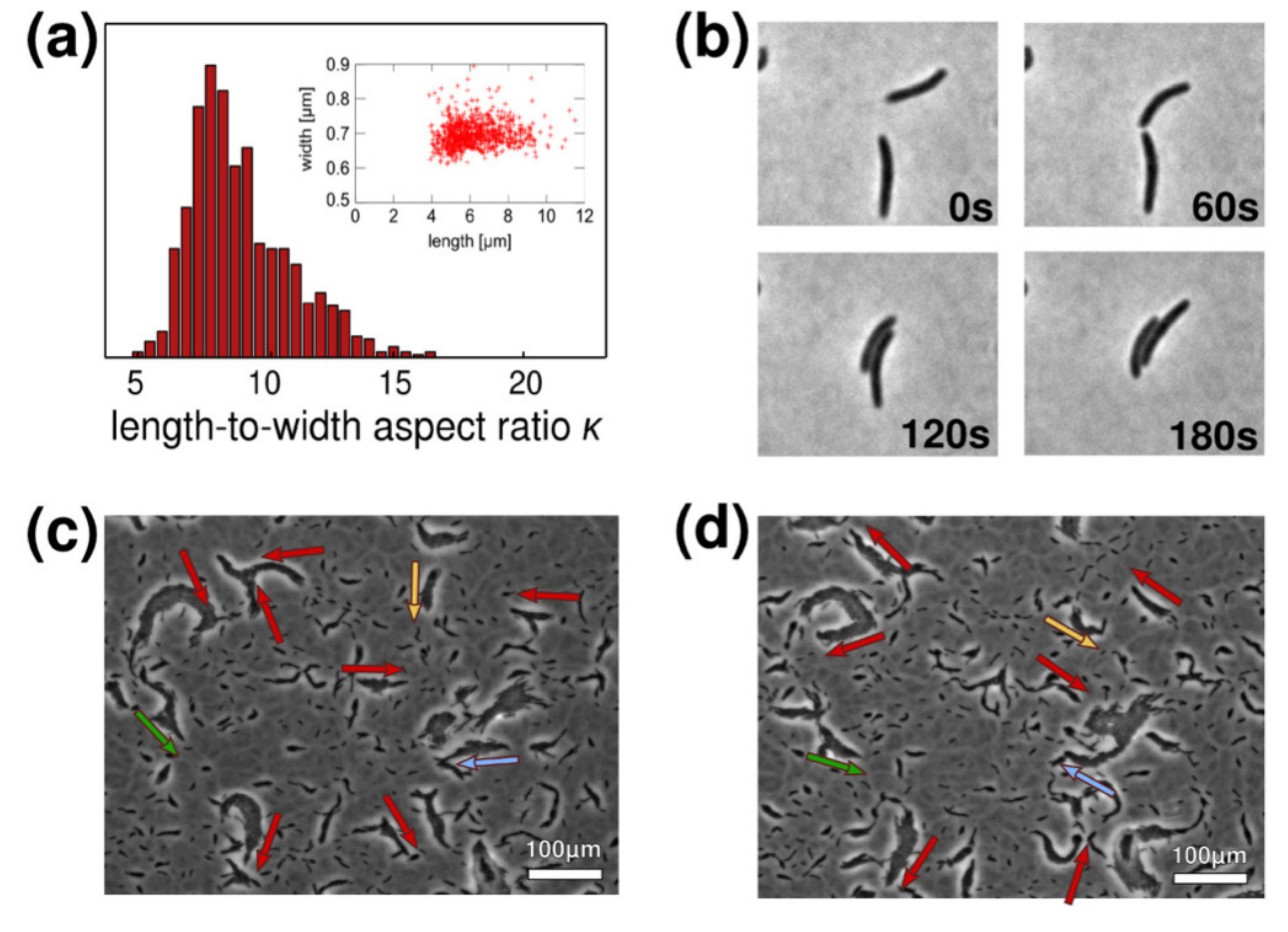}}}
\caption{%
Gliding bacteria ({\it Myxococcus xanthus})  are a good example of real-world active Brownian rods. 
The figure shows that individual collisions among the moving bacteria often lead to an effective alignment 
that in turn lead to the formation of moving polar clusters. The cluster size distribution (as function of the density) in myxobacterial experiments is remarkably similar to the one observed 
in simulations with active Brownian rods. Figure adapted from~\cite{peruani2012}.  
}
\label{fig:myxo}
\end{center}
\end{figure}

If elongated active, {\it i.e.} self-propelled, objects interact by  pushing each other in a dissipative medium or substrate, the objects will tend to locally align~\cite{peruani2006} 
as shown in Fig.~\ref{fig-model}.  
Since these object are self-propelled, once aligned, they will move together in the same direction for a given time. 
This simple effective alignment mechanism among active objects lead to interesting collective effects~\cite{peruani2006,yang2010,baskaran2008}, 
as the formation of moving cluster as illustrated in Fig.~\ref{fig:myxo} with experiments of myxobacteria.   
There is a broad range of real-world active systems that consist of active elongated object where this mechanism is at work: gliding bacteria~\cite{peruani2012,starruss2012},  dried self-propelled rods~\cite{kudrolli2008,kudrolli2010}, chemically-driven rods~\cite{paxton2004,mano2005}, and 
it has been recently argued that also -- neglecting hydrodynamic effects over steric effects -- in swimming bacteria~\cite{wensink2012,dunkel2013,zhang2010} and motility assays~\cite{schaller2010,sumino2012}. 

Here, we review the large-scale properties of collections of active Brownian elongated objects, in particular rods, moving in a dissipative medium/substrate. 
We address the problem by presenting three different models of decreasing complexity, which we refer to as model I, II, and III, respectively.  
Model I is the full physical active Brownian rod model introduced in~\cite{peruani2006} where particles 
exhibit a well-defined shape, possess an active force acting along the longest axis of the rod, and interact -- via volume exclusion effects -- by pushing each other.  
In model I there exists  a coupling of local density, orientational order, and speed, known to lead to density instabilities and collective phenomena in other active models~\cite{peruani2011,marchetti2012b,grossmann2012}. 
More importantly, in model I active stresses coexist with an an effective local alignment mechanism. 
Due to the combined effect of these two elements, model I displays exciting new physics unseen in other active models, such 
as the formation of highly dynamical aggregates that constantly eject giant polar cluster containing thousands of active rods~\cite{weitz2015}. 

If we remove from model I the active force, we end up with an equilibrium system (if noise terms have been adequately chosen). 
With the elongated rods interacting through steric repulsive forces,  
Onsager's argument on thin rods applies~\cite{onsager1949} and the system exhibits local nematic order above a given critical density. 
We discuss the possibility of local nematic order and quasi-long-ranged order (QLRO) in two-dimensions by introducing model II, 
which is a simplified version of model I without anactive force. 
Model II  allows us to argue that the symmetry of the interaction potential in model I is nematic. 
We introduce model III to show that the peculiar large-scale properties displayed by model I  do not result, as has been argued,   
from the combined effect of self-propulsion and an effective nematic velocity alignment mechanism. 
Model III is an active version of model II and a simplified version of model I without volume exclusion interactions. 
Let us recall that hat most flocking models assume a velocity alignment mechanism whose  symmetry is ferromagnetic~\cite{vicsek2012,marchetti2013}.
From model III, we learn that active particles with a nematic velocity alignment exhibit macroscopic nematic structures~\cite{ginelli2010}, which are not present in model I, 
which displays polar order at short scales and highly dynamical, highly fluctuating phase-separated phase~\cite{weitz2015}.  

Comparing model I, II, and III we disentangle the role of activity and interactions and identify the contribution of every modeling element. 
In particular, we find that by ignoring volume exclusion effects, local and global nematic order seems to be possible,  
while by adding steric interactions the system is dominated by the interplay of active stresses and local alignment, which prevents the formation of orientational order at large scales in two-dimensions. 

The physics of active elongated objects, despite its ubiquity in experimental systems,  remains still poorly understood. Here, we present a detailed state of the art 
of the unique collective properties of this fascinating physical system. 

\section{Models}  \label{sec:models}

\subsection{The full physical active Brownian rod model -- model I}  \label{sec:model_full}

Let us consider  $N$ active Brownian rods (ABR)  moving in a two-dimensional space of linear size $L$ with periodic boundary conditions. 
Each rod is driven by an active stress/force $F$ that is applied along the long axis of the particle. Interactions among rods are modeled through 
a repulsive potential, which we denote, for the $i$-th particle, by  $U_i$. 
The substrate  where the rods move acts as a momentum sink.  
There are three friction drag coefficients,  $\zeta_{\parallel}$, $\zeta_{\perp}$, and $\zeta_{\theta}$, 
which correspond to the drags experienced by the rods as the rod moves along the long axis, perpendicular to it, or as it rotates, respectively.
%
In the over-damped limit, the equations of motion of the $i$-th 
rod are given, as in~\cite{peruani2006}, by:
\begin{eqnarray}
\label{eq:evol_x}
\dot{\mathbf{x}}_i &=&  \boldsymbol{\mu}  \left[ -\boldsymbol{\nabla}U_i +  F \mathbf{V}(\theta_i)+  \boldsymbol{\sigma}_i(t) \right]  \\
\label{eq:evol_theta}
\dot{\theta}_i       &=&    \frac{1}{\zeta_{\theta}}  \left[  - \frac{\partial U_i}{\partial \theta_i} +   \xi_{i}(t) \right]  \, , 
\end{eqnarray}
where the dot denotes a temporal derivative, $\mathbf{x}_i$ corresponds to the position of the center of mass and $\theta_i$ the orientation of the long axis of the rod. 
The term $U_i$ models the interactions with other rods and $F$ is the self-propelling force. 
The symbol $\boldsymbol{\mu}$ in Eq.~(\ref{eq:evol_x})  is the mobility tensor defined as  $\boldsymbol{\mu} = \zeta_{\parallel}^{-1} \mathbf{V}(\theta_i) \mathbf{V}(\theta_i) + \zeta_{\perp}^{-1} \mathbf{V}_{\perp} (\theta_i) \mathbf{V}_{\perp}(\theta_i)$, 
with $\mathbf{V}(\theta)\equiv (\cos(\theta),\sin(\theta))$ and $\mathbf{V}_{\perp}(\theta)$ such that  $\mathbf{V}(\theta).\mathbf{V}_{\perp}(\theta)=0$.
%
%
%
Drag friction coefficients can be computed assuming that the rods are surrounded by a liquid~\cite{doi1986,levine2004}, 
move on a dried surface as in experiments with granular rods~\cite{kudrolli2008}, or by assuming that Eqs. (\ref{eq:evol_x}) and (\ref{eq:evol_theta})  
represent gliding bacteria, in which case the friction coefficients are arguably connected to presence of the so-called focal adhesion points~\cite{peruani2012}. 
In short, the friction coefficients depend on the specific rod system Eqs. (\ref{eq:evol_x}) and (\ref{eq:evol_theta}) are supposed to model. 
Here, we use $\zeta_\parallel=10$, $\zeta_\perp=25$, $\zeta_\theta=2$, and $F=0.4$. 
Eq.~(\ref{eq:evol_theta}) represents the  temporal evolution of the orientation of the rod, which is assumed 
to result from the torque generated by the interactions with the others rods, modeled by $U_i$, and thus we express this torque as $-\frac{\partial U_i}{\partial \theta_i}$. 
Note that Eqs. (\ref{eq:evol_x}) and~(\ref{eq:evol_theta}) are subject to fluctuations through 
the terms $\boldsymbol{\sigma}_i(t)$ and $\xi_{i}(t)$, which correspond to  delta-correlated vectorial and scalar noise, respectively. 
For simplicity, we neglect  $\boldsymbol{\sigma}_i(t)$ and specify in  Eq.~(\ref{eq:evol_theta})  $\langle \xi_{i}(t) \rangle = 0$ 
and $\langle \xi_{i}(t) \xi_{j}(t') \rangle = 2 D_{\theta} \delta_{i,j} \delta(t-t')$, with $D_{\theta}=2.52\times10^{-2}$ (for more details see~\cite{weitz2015}).  
%
%
%
The interactions among the rods are modeled by a soft-core potential that penalizes particle overlapping. For the $i$-th rod, the potential takes the form:   
$U_i = U(\mathbf{x}_i, \theta_i) = \sum_{j=1;j \neq i}^{N} u_{i,j}$ ,
%
where $u_{i,j}$ denotes the repulsive potential interaction between the $i$-th and $j$-th rod, both of length $\ell$ and width $w$, 
such that  $u_{i,j}=u(\mathbf{x}_i-\mathbf{x}_j, \theta_i-\theta_j)$.
The rods (in two-dimensions) can be represented as  soft rectangles or spherocylinders as in~\cite{peruani2006} or  equivalently as  straight chains of $n$ disks of diameter $w$, as implemented in~\cite{wensink2012b,abkenar2013,mccandlish2012}, whose centers are separated a given distance $\Delta$; here $\Delta=w/3$. 
We notice that results obtained with active Brownian rods represented by disk-chains are qualitatively identical to those produced with the original ABR model introduced in~\cite{peruani2006}   
if and only if $\Delta \ll 2w$, {\it i.e.} as long as the border of the rods is smooth.   
Using this implementation,  $u_{i,j}$ can be  expressed as  $u_{i,j}=\sum_{\alpha, \beta} u_{i,j}^{\alpha, \beta}$, where 
$u_{i,j}^{\alpha, \beta}$ is the potential between disk $\alpha$ of the $i$-th rod and disk $\beta$ of the $j$-th rod, which here we assume to be given by a harmonic repulsive potential:  
$u_{i,j}^{\alpha, \beta}=C_0\left(d^{i,j}_{\alpha, \beta}-w\right)^2$, for $d^{i,j}_{\alpha, \beta}<w$ and zero otherwise, 
where $d^{i,j}_{\alpha, \beta}$ is the distance between the centers of the disks and $C_0=200$.

%

\subsection{The (passive) Brownian rod model -- model II}  \label{sec:model_passive}

If we remove the active force and make $F=0$ in Eq.~(\ref{eq:evol_x}), while requiring that $\boldsymbol{\sigma}_i(t)$ and $\xi_{i}(t)$ 
allow us to define a temperature, the resulting system is the equilibrium system of passive rods studied by Onsager~\cite{onsager1949} 
to describe, in a simple way, rod-like liquid crystals in two-dimensions. 
Let us recall that  according to the so-called Mermin-Wagner theorem~\cite{mermin1966}, this equilibrium system in two-dimensions cannot exhibit long-ranged order (LRO). 
Furthermore, it is believed that this system exhibits (in two-dimensions) a defect-mediated phase transition analogous to the  
Berezinskii-Kosterlitz-Thouless (BKT) transition~\cite{berezinskii_destruction_1971,kosterlitz_ordering_1973}. 
This means that above a given critical point,  the system exhibits quasi-long-ranged order (QLRO), implying that by increasing the system size, while keeping all intensive parameter fixed, the order parameter (associated to orientational order) decreases algebraically~\footnote{Some recent numerical works~\cite{farinas2010} using the Lebwohl-Lasher model~\cite{lebwohl1973} have called into question 
the presence of QLRO in two-dimension in such model  (in three-dimensions, the transition is believed to be of weak first-order).}. 

At the mean-field level, {\it i.e.} by neglecting fluctuations, the equilibrium model II displays an isotropic-nematic transition, which can be understood by considering two rods separated (their center of mass) a given distance $d<L$. 
To simplify the reasoning, let us neglect the dynamics of the center of mass of the rods and consider them fixed.  
In this scenario, it is clear that if the interaction between the rods is modeled by an interaction potential that penalizes the overlapping of the rods (as in model I), 
by making the orientation of the rods parallel to each other, we minimize the interaction potential (moreover, particles may not even interact). 
In addition, since the interaction potential does not distinguish head and tail of the particles, it should be such that we obtain the same by 
flipping the orientation of one of the particles by $\theta_i \to \theta_i + \pi$. 
In short, the interaction potential has to be a function of overlapping area of the two rods, which is, as observed by Onsager~\cite{onsager1949}, proportional to $-\cos^2(\theta_j - \theta_i)$, where $i$ and $j$ refer to the label of the particles we are looking at. 
Since $\cos^2(u)=\left(1+\cos(2 u)\right)/2$, for simplicity 
we express the potential directly as: 
\begin{eqnarray}
\label{eq:pot_model_II}
U_i(\mathbf{x}_i, \theta_i) = -\frac{\hat{\gamma}}{2} \sum_{|\mathbf{x}_i - \mathbf{x}_j|<R} \cos \left(2(\theta_j - \theta_i)  \right) \, ,
\end{eqnarray}
where the sum runs over all $j$ particles such that $|\mathbf{x}_i - \mathbf{x}_j|<R$, with $R$ defining the interaction range, which we could assume to be $R \sim \ell$, 
and $\hat{\gamma}$ a constant.  
This defines a simplified dynamics where the equation of motion of the $i$-th particle is given by:
\begin{eqnarray}
\label{eq:evol_x_model_II}
\dot{\mathbf{x}}_i &=&    \hat{\boldsymbol{\sigma}}_i(t)   \\
\label{eq:evol_theta_model_II}
\dot{\theta}_i       &=&  -\frac{1}{\zeta_{\theta}} \frac{\partial U_i}{\partial \theta_i} +   \xi_{i}(t)  =  \gamma \sum_{|\mathbf{x}_i - \mathbf{x}_j|<R} \sin\left( 2 (\theta_j - \theta_i)\right)  +   \xi_{i}(t)  \, ,  
\end{eqnarray}
where $\gamma = \hat{\gamma}/\zeta_{\theta}$ and where we have simplified further the model by assuming that $\hat{\boldsymbol{\sigma}}_i(t)$ is  an isotropic delta-correlated vectorial noise.  
It is important to stress that Eqs. (\ref{eq:evol_x_model_II}) and (\ref{eq:evol_theta_model_II}) 
do not correspond (exactly) to Eqs. (\ref{eq:evol_x}) and (\ref{eq:evol_theta}) with $F=0$, 
but to a simplified version of model I for $F=0$ that shares the same symmetry. 
The fundamental difference between model I with $F=0$ and model II is the absence of a term proportional to $-\nabla U_i$ in Eq. (\ref{eq:evol_x_model_II}). 
Despite of this difference, we expect that on large scales Eqs. (\ref{eq:evol_x_model_II}) and (\ref{eq:evol_theta_model_II}) to display a behavior 
analogous to that of model I with $F=0$. 
We know that a system evolving according to Eqs. (\ref{eq:evol_x_model_II}) and (\ref{eq:evol_theta_model_II}) 
do not exhibit LRO~\footnote{The analysis of this system, which strictly speaking is non-equilibrium, can be performed along the same line as in~\cite{robert}.}, and that the observed transition is a defect-mediated transition with order being QLRO as expected for model I with $F=0$. 
In addition, and very important for us, at the mean-field level Eqs. (\ref{eq:evol_x_model_II}) and (\ref{eq:evol_theta_model_II}) exhibit 
an isotropic-nematic transition, characterized by the absence of local and global polar order. 

We will use model II as a reference model. The message is that in the absence of activity (meaning for $F=0$) and in two-dimesions, we 
do not expect the system to exhibit LRO but rather QLRO. Furthermore, from model II we could imagine that if activity induces order at large scales ({\it i.e.} LRO) 
such order should be nematic. We will see that this is the case in model III, which we introduce next, but counterintuitive not for model I.

\subsection{The idealized active Brownian rod model -- model III}  \label{sec:model_idelized}

The third and final model we introduce here is an active version of  model II~\cite{peruani2008}. 
The equations of motion of the $i$-th particle are given by: 
\begin{eqnarray}
\label{eq:evol_x_model_III}
\dot{\mathbf{x}}_i &=& v_0   \mathbf{V}(\theta_i)   \\
\label{eq:evol_theta_model_III}
\dot{\theta}_i       &=&    \gamma \sum_{|\mathbf{x}_i - \mathbf{x}_j|<R} \sin\left( 2 (\theta_j - \theta_i)\right)  +   \xi_{i}(t)  \, . 
\end{eqnarray}
where $v_0 = F/\zeta_{\parallel}$. 
Despite the fact that the interaction potentials in model I and model III are short-ranged and  
share the same symmetry, there is a fundamental difference between both models:  
the equation for $\dot{\mathbf{x}}_i$ in model III does not possess a term $-\nabla U_i$. 
The absence of this term implies that in model III there is no volume exclusion, and thus  
particles are point-like. 
As it will become clear below, this leads to the absence of active stresses in model III, which are present in model I. 
Due to this  difference, which may appear at first glance minor, model I and III exhibit qualitatively different large-scale properties. 
Another difference between model I and III is the absence of a noise terms in Eq. (\ref{eq:evol_x_model_III}), while such terms are present in Eq. (\ref{eq:evol_x}). 
Such noise terms do not have an impact on the (qualitative) large-scale properties of the systems. 

Finally, the difference between model II and III is given by the so-called active term, $v_0   \mathbf{V}(\theta_i)$, in Eq. (\ref{eq:evol_x_model_III}), 
replacing the noise term in Eq. (\ref{eq:evol_x_model_II}). 
Note that in model III the equations for $\dot{\mathbf{x}}_i$ and $\dot{\theta}_i $ are thus coupled through $\mathbf{V}(\theta_i)$, while this does not occur in model II. 
We will see below that the presence of the active term in model III has a strong impact on the large-scale properties, which turn to be qualitatively different from 
those of model II. 

\section{Order parameters and other observables -- definitions}

We characterize the resulting macroscopic patterns by their level of orientational order through 
\begin{eqnarray}
\label{eq:op_order}
 S_q   = \langle S_q(t) \rangle_t = \langle | \langle \exp(\imath \,q\,\theta_i(t)) \rangle_{i} | \rangle_t \, ,
\end{eqnarray}
where the averages run over the number of particles and time. Polar order corresponds to $q=1$ and  nematic order to $q=2$. 
The $S_1$ and $S_2$ correspond to global order parameters, {\it i.e.} obtained by performing an average over the entire system. 
Local order parameters can be also defined and should be noticed that 
a system can exhibit no global order, while still displaying local order. 

The distribution of particles in space as well as information on the interaction structure can be studied 
by looking at the cluster size distribution (CSD). 
We use the (weighted steady state) CSD $p(m)$ defined as the time average (after an initial transient) of the instantaneous CSD: 
\begin{eqnarray}
p(m,t)=\frac{m~n_m(t)}{N},
\end{eqnarray}
where $n_m(t)$ is the number of clusters of mass $m$ present in the system at time $t$.
Notice that the normalization of this distribution is ensured since $N=\sum_{i=1}^N m~n_m(t)$ (with $N$ the number of particles in the system as defined above).  
Clusters are collections of interconnected particles, where any two particles are considered as (directly) connected if they interact 
to each other. 
For simplicity and to speed up the numerics, the criterion is usually relaxed and particles are considered as connected if they are 
separated a given distance which is typically of the order of the interaction radius (the details vary from model to model). 
The functional form of $p(m)$ indicates whether the system is ``homogeneous'', when the CSD displays an exponential tail, or 
phase-separated, when the CSD exhibits a peak at large cluster sizes. 
Particles can also self-organized in a kind of ``critical'' state where cluster of all sizes can be observed. This corresponds to a heavy-tailed CSD. 
A detailed analysis of the CSD of these models can be found in~\cite{peruani2006,peruani2013,peruani2010}.  
An alternative way to monitor phase separation is by studying the ratio between average cluster size $\langle m \rangle$ and system size $N$, where $\langle m \rangle =  \sum_m m\, p(m)$, with the advantage of dealing simply with an scalar, $\phi = \langle m \rangle/N$. 

\begin{figure}[t]
\begin{center}
\includegraphics[clip,width=16cm]{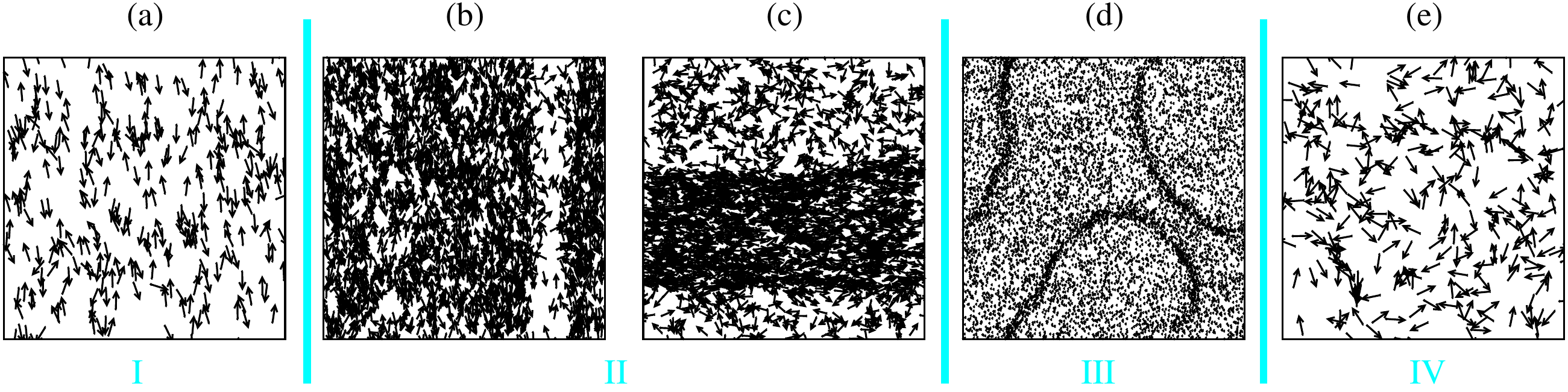}
\caption{[model III] (a-c) Stationary state
typical snapshots at different noise values $\Gamma$ 
($\rho=1/8$ $v_{\rm 0}=0.5$ and $L=2048$)
The arrows indicate particles direction of motion; only a fraction of 
total particles are shown for clarity reasons.
(a) $\eta=0.08$, (b) $\eta=0.10$, (c) $\eta=0.13$, (d) $\eta=0.168$
(for clarity reasons particles are represented only by points),
(e) $\eta=0.20$.
Roman numbers refers to the four phases observed in finite size systems. Figure from~\cite{ginelli2010}.}
\label{fig:2}
\end{center}
\end{figure}

In order to study the thermodynamical behavior of the models, we analyze the scaling  
of $S_q$ and $\phi$ with the system size $N$, while keeping all intensive parameter fixed. 
If $S_q \propto C_{\infty} + f(N)$, with $f(N\to\infty)\to0$, the system exhibits LRO. 
QLRO, on the other hand, corresponds to  $S_q \propto N^{-\beta}$, where $\beta \leq \beta_0$ 
with $\beta_0 \sim 1/16$ in two-dimensions for the XY model. 
It is worth recalling that for a fully disorder system, we expect $S_q \propto N^{-1/2}$, which means 
that observing an algebraic decay does not necessarily imply QLRO. 
Finally, $\phi \propto N^{-1}$ for homogeneous system, indicating that there is a finite value of  $\langle m \rangle$ with $N$. 
A departure from this behavior is an indication of either phase-separation or the organization of particles in a ``critical" CSD.

\section{Large-scale properties of the models}  \label{sec:results_models}

Here, we review results for model I and III. 
The large-scale properties of model II have been already summarized in the model section. 
As mentioned before, model II, {\it i.e.} the Brownian rod model, is used as a reference model in order 
to understand the role of the active term. 
The comparison between model I and III will allow us to disentangle the role of volume-exclusion and nematic alignment. 
We start in inverse order, that is, we first revise results for model III, since the model is simpler and results are, somehow, 
more intuitive.  
\begin{figure}
\begin{center}
\includegraphics[clip,width=10cm]{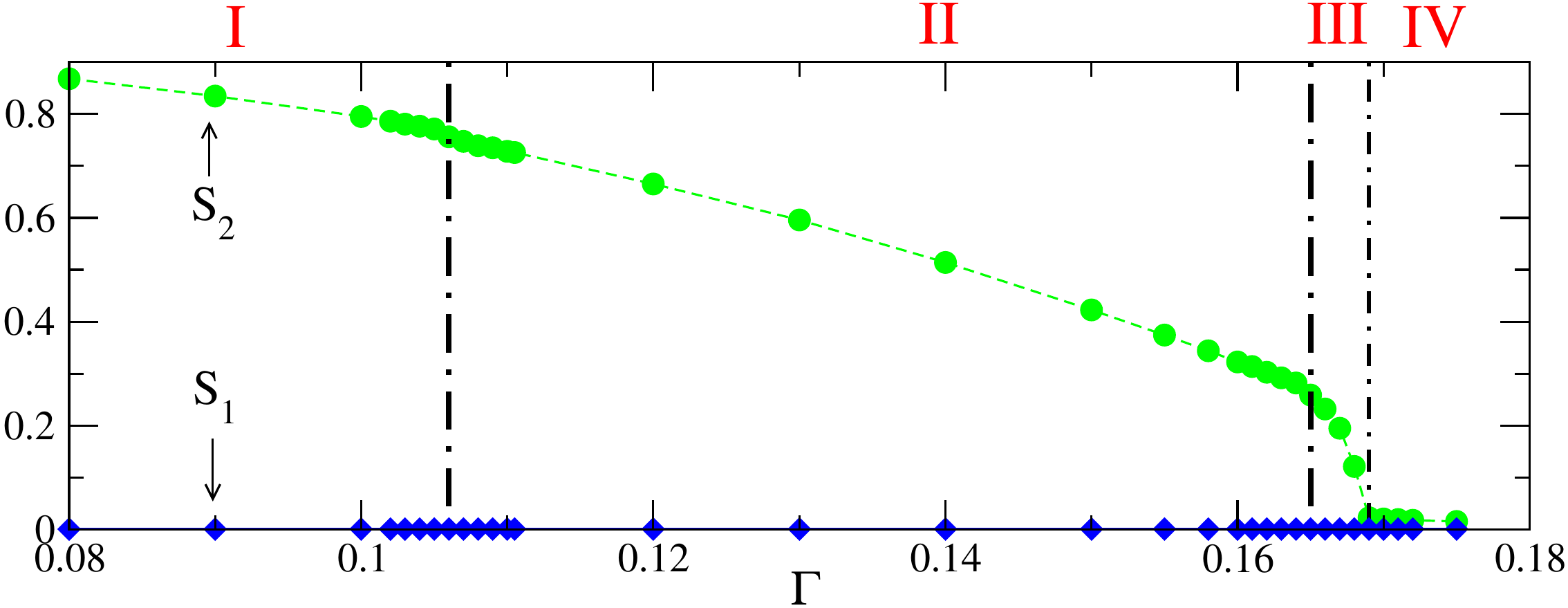}
\caption{[model III] Nematic order parameter $S_2$ as 
function of the  noise amplitude $\Gamma$ for a density $\rho=1/8$ and a 
linear system size $L=2048$. The figure illustrates the four different regimes 
observed for a given density and system size.
}
\label{fig:3}
\end{center}
\end{figure}

\subsection{Nematic order and band formation in the idealized rod model -- model III}  \label{sec:result_model_III}

The numerical results that we are going to review~\cite{ginelli2010} here correspond to 
an implementation of model III {\it {\`a} la Vicsek}, {\it i.e.} using a numerical scheme as in~\cite{vicsek1995}. 
It is important to stress that direct numerical interaction of the continuum-time Eqs.~(\ref{eq:evol_x_model_III}) and~(\ref{eq:evol_theta_model_III}) 
lead to the same qualitative behavior as the one reported in~\cite{ginelli2010} (data not shown). 
Results where generated by using the following specific scheme~\footnote{
Simulations results are undistinguishable if Eq.(\ref{motion_pos}) is 
replaced with $\mathbf{x}_j^{t+1}= \mathbf{x}_j^{t}+v_0\, \mathbf{V}(\theta_k^{t+1})$, which is the actual implementation used in~\cite{ginelli2010}.} :
\begin{eqnarray}
\mathbf{x}_j^{t+1}&=& \mathbf{x}_j^{t}+v_0\, \mathbf{V}(\theta_k^{t})
\label{motion_pos} \\
\label{motion_angle}
\theta_j^{t+1}&=& 
\arg \left[\sum_{k\sim j}{\rm sign}\left[\cos(\theta_k^t -\theta_j^t)\right] 
\mathbf{V}(\theta_i^t )\right] + \Gamma \psi_{j}^{t} \, ,
\end{eqnarray}
where $\arg[\mathbf{b}]$ refers to the angle associated to the vector $\mathbf{b}$ if this is expressed in polar coordinates, 
the sum is   taken over all particles $k$ within distance $1$ of $j$ (including $j$ itself) 
and $\psi_{j}^{t}$ is a white noise uniformly distributed in 
$\left[-\frac{\pi}{2},\frac{\pi}{2}\right]$. The term $\Gamma$ is the noise amplitude, which means that  
$D_{\theta} \propto \Gamma^2$. 
Simulations were performed with $v_0=1/2$ and particle density $\rho=N/L^2=1/8$. 

A fluctuating ordered phase exists
at low noise (or high density if $\rho$ is used as control parameter), and 
an order/disorder transition line lies in 
the main parameter plane $(\rho,\Gamma)$. 
Varying $\Gamma$, we observe in a square domain of linear size $L=2048$, as shown in Fig.~\ref{fig:2} and \ref{fig:3}, 
that only {\it nematic} orientational order arises. Despite the polar nature
of the particles, the polar order parameter remains near zero for all
noise strengths (Fig.~\ref{fig:3}). 
Both the ordered and the disordered phases are 
divided in two by the spontaneous segregation, at intermediate $\Gamma$ values,
of the system into high-density, ordered regions and sparse, disordered ones
(Fig.~\ref{fig:2}b-d). A total of four phases thus seems to exist, 
labeled I to IV by 
increasing noise strength hereafter. Phases I and II are nematically ordered,
phases III and IV are disordered.

\begin{figure} 
\begin{center}
\includegraphics[clip,width=8.0cm]{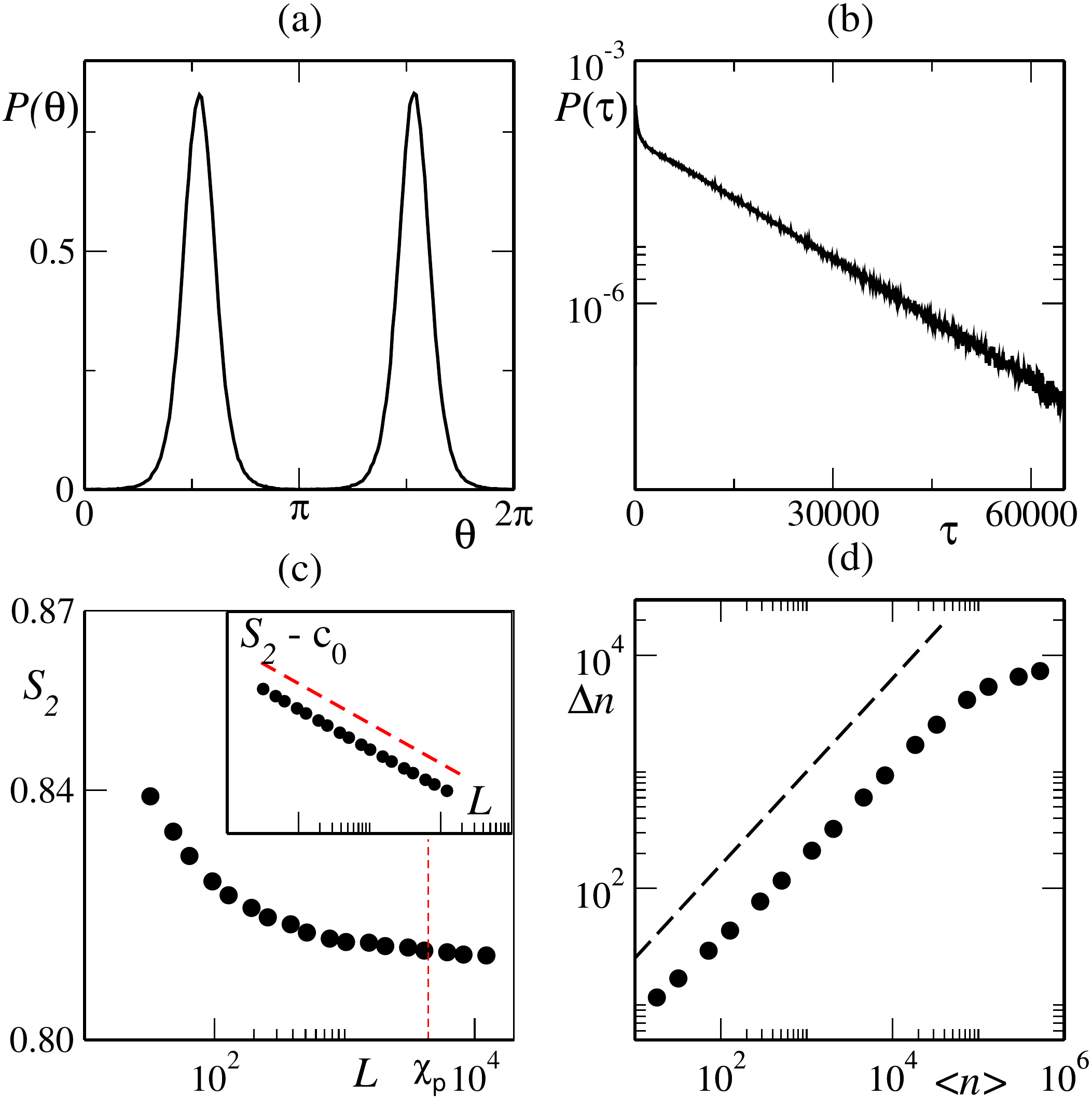}
\caption{[model III]  
Ordered homogeneous phase ($\rho = 1/8$, $v_{\rm0}=0.5$, $\eta=0.095$).
(a) Moving direction (given by $\theta$)  distribution in a system of linear system size $L=2048$.  
(b) Distribution of particles transition times $\tau$ between the two peaks of panel (a), i.e. particles
which ``turn around'' their orientation by  $\approx \pi$. 
(c) Nematic order parameter $S_2$ as a function 
of system size $L$. The vertical red dashed line marks the persistence length 
$\chi_{\rm f} \approx 4400$. 
In the inset: A constant value $c_{\infty}=0.813063$ has been subtracted from the nematic order parameter
to highlight power low decay to a nonzero constant. The red dashed line decays as $L^{-2/3}$.
(d) Number fluctuations $\Delta n$ as a function of average particle number
$\langle n \rangle$  in a system with $L=4096$. 
The dashed line marks the power law growth $\langle n \rangle^{0.8}$. Figure taken from~\cite{ginelli2010}.}
\label{fig:4}
\end{center}
\end{figure}

Phase I, present at the lowest $\Gamma$ values, is ordered and 
spatially homogeneous (Fig.~\ref{fig:2}a). Its nematic order,
which arises quickly from any initial condition, is the
superposition, at any time, of two polarly aligned  opposite subpopulations of
statistically equal size (Fig.~\ref{fig:4}a). These subpopulations
constantly exchange particles, those which ``turn around'', an event 
that occurs on exponentially-distributed times $\tau$ (Fig.~\ref{fig:4}b). Therefore it is natural to define a 
typical persistance time $\tau_{\rm p}$ and its corresponding persistence length 
$\chi_{\rm p} \approx v_0 \tau_{\rm p}$.
They are found to decrease as the noise amplitude is increased.
While the breaking of a continuous symmetry in a two dimensional equilibrium system 
can only lead to quasi-long-range-order (QLRO) as discussed for model II, 
in model III the numerical evidence points towards LRO. 
Let us recall that what has been called as ``active nematics"~\cite{chate2006}, 
which is equivalent to model II, where we have to explicitly require the use of  a 
local diffusion tensor depending on $\theta$, displays QLRO. 
Here the {\it nematic} order parameter $S_2$, for the tested system sizes, decays slower 
than a power law. A good fit of this decay is given
by an algebraic approach to a constant asymptotic value $c_{\infty}$ 
(Fig.~\ref{fig:4}c). This phase is characterized by the presence of giant number fluctuations, Fig.~\ref{fig:4}c (for details see~\cite{ginelli2010}). In short, the numerical data suggests the existence of
true long-range nematic order.

Phase II differs from phase I by the presence, in the steady-state,
of a low-density disordered region. In large-enough systems,
a narrow, low density channel emerges (Fig.~\ref{fig:2}b)
when increasing $\Gamma$. It becomes wider and wider at larger $\Gamma$ values, 
so that one can then speak of a high-density nematically ordered  band amidst a 
disordered sea (Fig.~\ref{fig:2}c). 
Inside the band, nematic order similar to phase I is found, with 
particles traveling in roughly equal number along each direction, and turning
around or leaving the band at exponentially-distributed times, albeit with a shorter typical persistence time. 
Giant number fluctuations similar to those reported in Fig.~\ref{fig:4}d occur.
In rectangular domains, the band is typically oriented along the small dimension of the box.
Bands along the longer dimension can be artificially created, 
but in sufficiently long and narrow boxes they become wavy under what looks like
a finite-wavelength instability and are eventually destroyed, 
leaving a thicker band along the small dimension.
For large-enough (square) domains, the  band possesses
a well-defined profile with sharper and sharper edges as $L$ increases.

In phase III, spontaneous segregation into bands still occurs 
(for large-enough domains), but these now thin structures constantly bend and 
elongate, 
getting thinner and vanishing, or merging with others. 
They never reach a static steady state shape. 
Correspondingly, the nematic order parameter fluctuates on very large time
scales and $S_2$ decrease like $N^{-1/2}$ providing a clear indication that  
the order in these thin highly dynamical bands self-averages, making phase III a disordered
phase albeit one with huge correlation lengths and times.

Finally, Phase IV, observed for the highest noise strengths, is disordered on small
length- and time-scales, without any particular emerging structure (Fig.~\ref{fig:2}e).

\begin{figure}
\begin{center}
\resizebox{11cm}{!} {\includegraphics{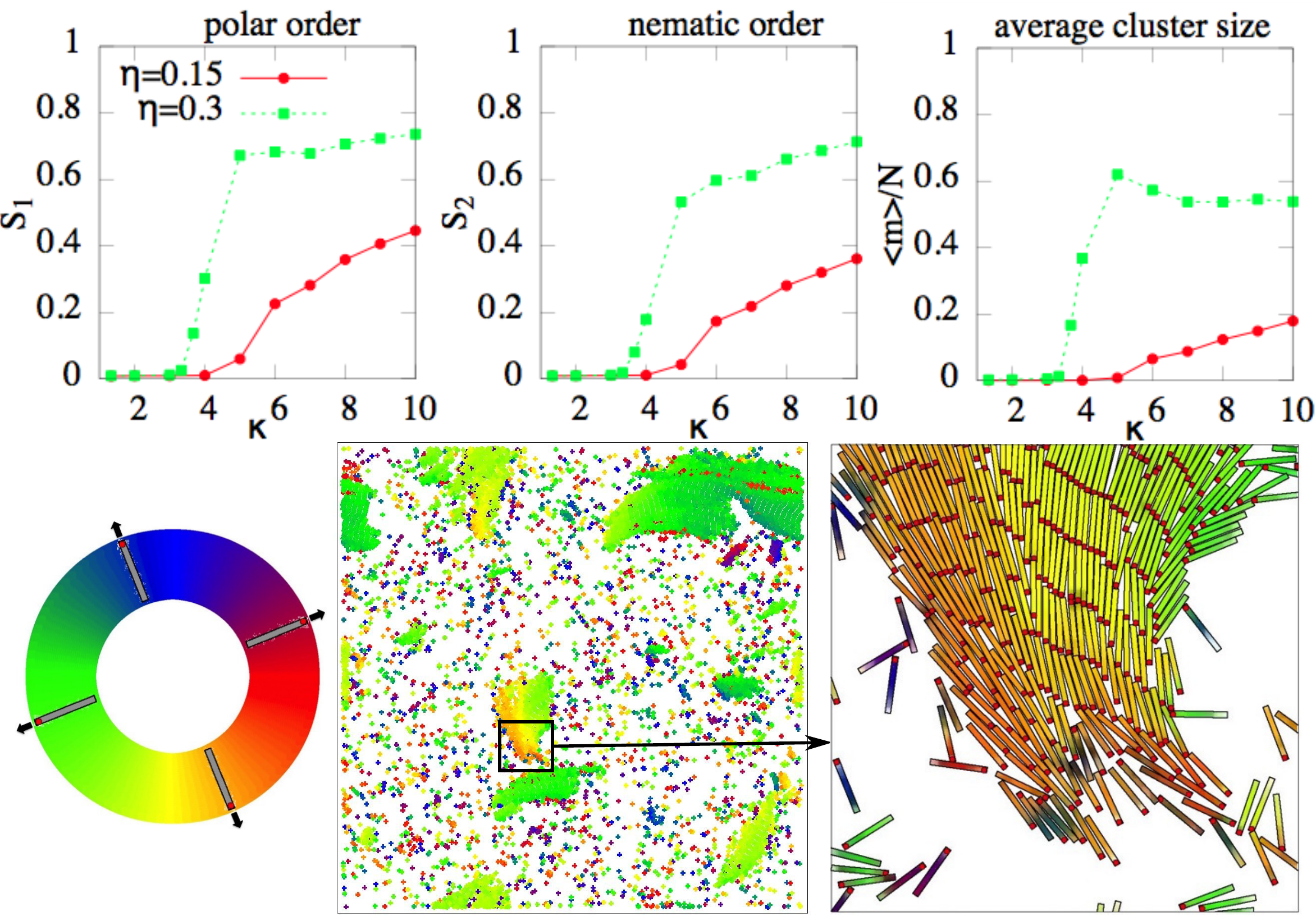}}
\caption{[model I] Transition from disorder to order and phase separation at small system sizes. Top, from left to right: polar order parameter $S_1$, nematic order parameter $S_2$, and average cluster size $\langle m \rangle$ over system size $N$, respectively, as function of the particle aspect ratio $\kappa$ for various packing fractions $\eta$. $N=10000$. Bottom: A simulation snapshot for $\kappa=10$ and $\eta=0.3$, with $S_1=0.5$, $S_2=0.26$, $\langle m \rangle/N=0.14$. Notice that clusters are polar. The orientation of rods is color coded as indicated in the bottom left panel. From~\cite{weitz2015}. 
 }
\label{fig:PT}
\end{center}
\end{figure}

\subsection{Absence of global orientational order, local polar order, and dynamic aggregates in the physical active Brownian rod model -- model I}  \label{sec:result_model_I}

Here, we study the long-scale properties of model I by direct integration of Eqs. (\ref{eq:evol_x}) and (\ref{eq:evol_theta}). 
We use as control parameter  the aspect ratio  $\kappa$, but we could use instead the packing fraction $\eta = a \,N/L^2$ or the noise amplitude $D_{\theta}$. 
It is important to clarify that when we vary the aspect ratio  $\kappa$, we keep fixed the particle area $a = \ell \times w$. 
For a fixed system size $N$, 
we observe  $S_1$, $S_2$, and $\langle m \rangle/N$ take off above a critical $\kappa_c$ value as shown in Fig.~\ref{fig:PT}. We notice that $\kappa_c$ decreases when $\eta$ is increased.
Below $\kappa_c$, we observe a gas phase characterized by the absence of orientational order and an exponential cluster size distribution 
such that  $\langle m \rangle/N \propto N^{-1}$. 
For $\kappa>\kappa_c$ the system undergoes  a symmetry breaking as observed previously in~\cite{baskaran2008,wensink2012b,abkenar2013}. 
The emerging order is, however,  polar as evidenced by the behavior of $S_1$. 
We recall that in the presence of polar order, $S_2$  is slaved to $S_1$. 
%
The behavior of $\langle m \rangle/N$, right panel in Fig.~\ref{fig:PT}, indicates that  the system starts to spontaneously self-segregate for $\kappa>\kappa_c$. 
Here, we find that the onsets of orientational order and phase separation coincide and share the same critical point, as predicted using a simple 
kinetic model for the clustering process~\cite{peruani2013}: due to the effective velocity alignment large polar clusters emerge, which in turn lead to macroscopic polar order. 
Notice that in an equilibrium system of (hard) rods (i.e. $F=0$) for $\eta \leq 0.3$ and $1 \leq \kappa \leq10$, according to De las Heras et al.~\cite{delasheras2013}, 
we should observe only an homogeneous disordered phase for this range of parameters.  
This indicates that the observed phase transition requires $F>0$, i.e. the active motion of the rods.

\begin{figure} 
\begin{center}
\resizebox{10cm}{!} {\includegraphics{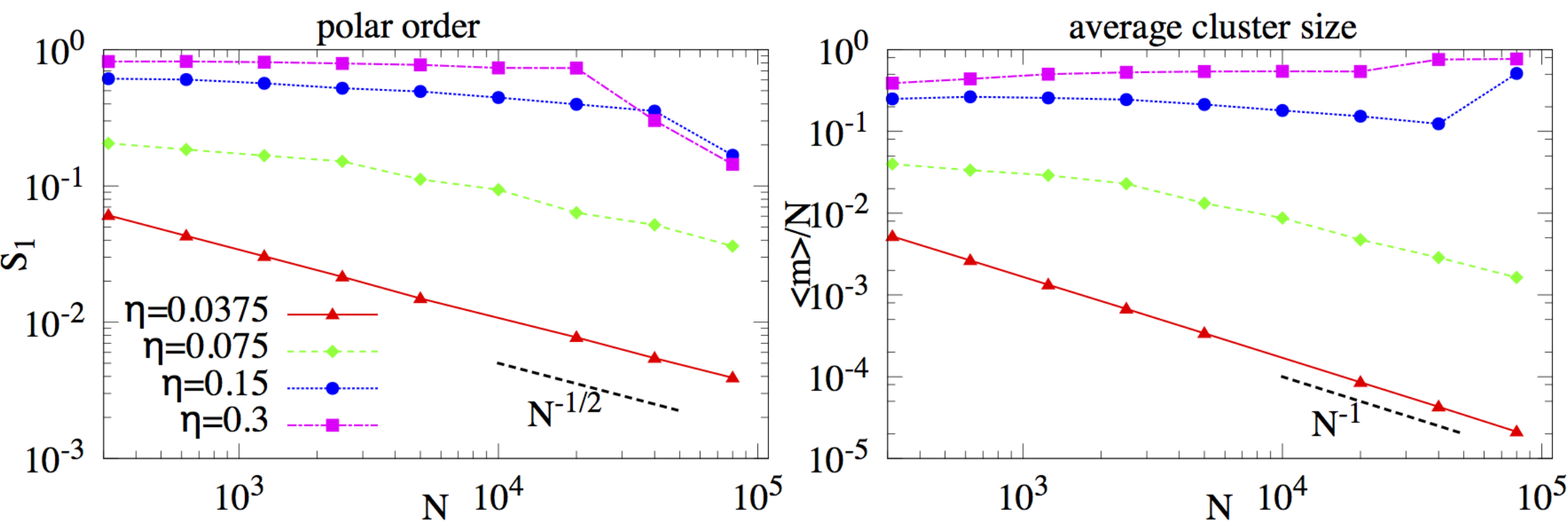}}
\caption{[model I] Finite size scaling of the polar order parameter $S_1$ and average cluster size with respect to system size $\langle m \rangle/N$ for aspect ratio $\kappa = 10$ and several packing fractions $\eta$. 
Notice that, for $\eta \geq 0.15$, the system becomes disordered as the system size $N$ is increased, while remaining phase-separated. From~\cite{weitz2015}.  
 }
\label{fig:FSS}
\end{center}
\end{figure}

We performed a finite size study, by increasing simultaneously $N$ and $L$ while keeping the packing fraction $\eta$ and all other parameters constant. Fig.~\ref{fig:FSS} shows the scaling of the (global) polar order parameter $S_1$ and average cluster size with respect to system size $\langle m \rangle/N$ for aspect ratio $\kappa = 10$ and several packing fractions $\eta$. 
At low $\eta$ values, i.e. for $\eta \leq 0.075$, we are in the situation $\kappa < \kappa_c$ (we recall that the critical $\kappa_c$ value depends on $\eta$). For $\kappa < \kappa_c$ the system is not phase-separated and does not exhibit orientational order. As expected, the scaling of $S_1$ with $N$ shows that  $S_1 \propto N^{-\alpha}$, with $\alpha = 1/2$, which means that the system is fully disordered. In addition, we observe that  $\frac{<m>}{N} \propto N^{-\beta}$, with $\beta = 1$, which indicates that there is a well-defined characteristic cluster size for the system (that is independent of $N$) and consequently the system is not phase-separated.
%

At large $\eta$ values, i.e. for $\eta \geq 0.15$ and $\kappa > \kappa_c$,  we observe phase separation and (global) polar order for small finite systems ($N=10000$). The finite size study shows that, for $\kappa > \kappa_c$, $\frac{<m>}{N}$ does not decrease (asymptotically) with $N$. Moreover, for large values of $N$, we even observe an increase. This indicates that $\langle m \rangle$ is at least proportional to $N$, which implies that the system is phase-separated in the thermodynamical limit as well as in finite systems.
At the level of the orientational order  parameter $S_1$, we observe an abrupt change in scaling of $S_1$ with $N$. 
For $N<N_*$ (e.g., $N_*\sim 20000$) for $\eta=0.3$ and $\kappa=10$, $S_1$ is high {\it i.e.} the system displays  global polar order . 
On the other hand,  for $N>N_*$, while the system remains phase-separated, $S_1$ sharply decreases with $N$. 
In short, the finite size study reveals that, although phase separation does take place in the thermodynamical limit as well as in finite systems, the phase transition to an orientationally ordered phase is observed only for small finite systems. 
%
Global order patterns are not present in the thermodynamical limit, with the phase transition occurring, in this limit, between a disordered gas and a phase-separated state with no global orientational order. 
 %
%

\begin{figure} 
\begin{center}
\resizebox{10cm}{!}{\rotatebox{0}{\includegraphics{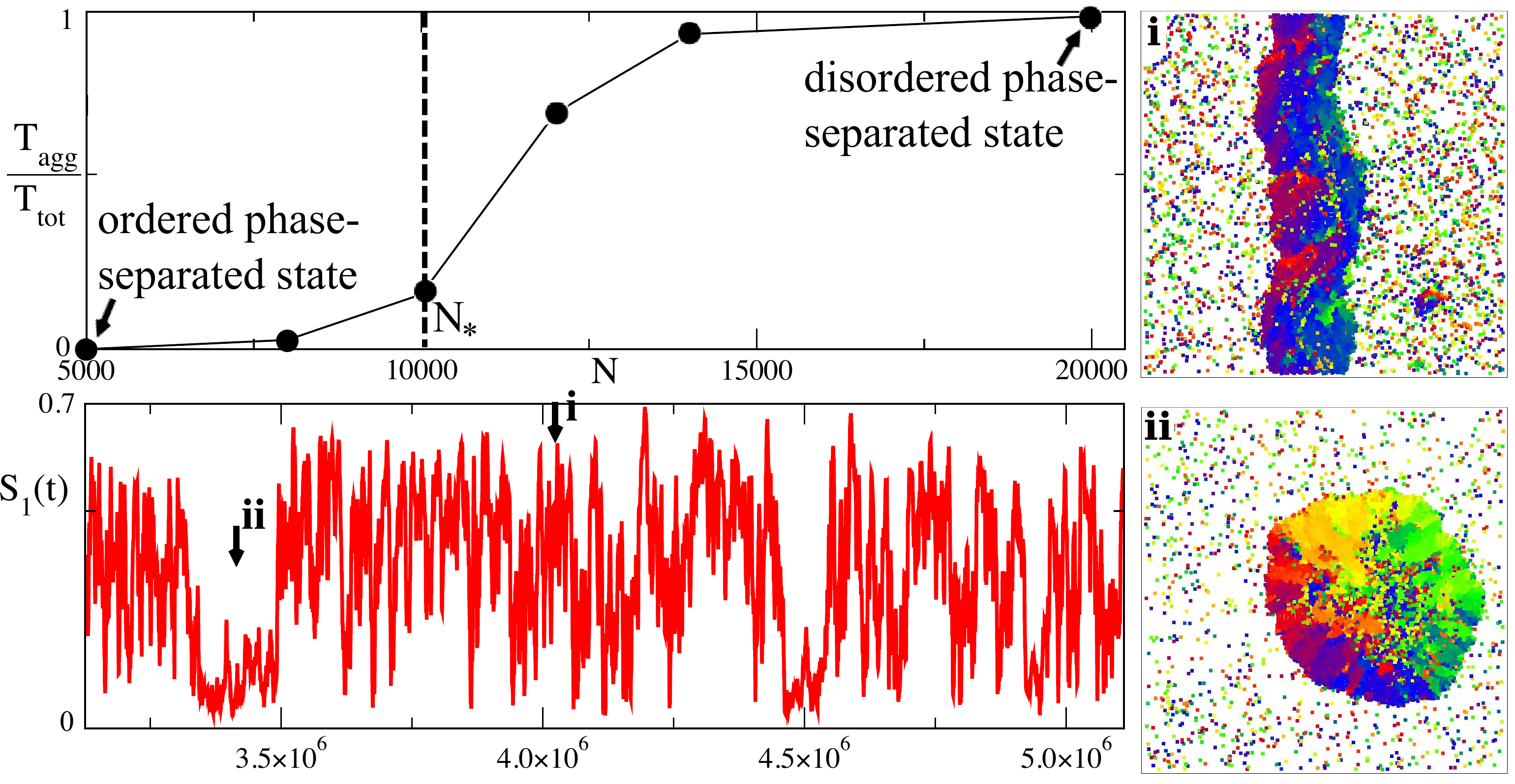}}}
\caption{ [model I] 
The orientationally ordered phase-separated state observed for $\kappa>\kappa_c$ in small systems  ($N\ll N_*$) becomes instable when $N$ is increased. In very big systems ($N\gg N_*$) we observe only aggregates (which correspond to an orientationally disordered phase-separated state).
Top left panel:   Evolution of  $T_{agg}/T_{tot}$ with $N$, where  $T_{agg}$ is the time the system spent in the aggregate phase and $T_{tot}$ -- here, $T_{tot}=10^7$ --  is the total simulation time.
Bottom left panel: the polar order $S_1$ as function of time for $N\sim N_*$ -- here, $N_*\sim 10^4$.
Large values of $S_1$ correspond to the system being highly ordered, typically due to the formation of a highly ordered band (panel i), 
while low values of   $S_1$ correspond to the formation of an aggregate (panel ii).  
Simulations correspond to $\kappa = 4$ and $\eta = 0.3$. Adapted from~\cite{weitz2015}. 
}
\label{fig:band}
\end{center}
\end{figure}

%
The reason for observing non-vanishing global order in small systems is the presence of few giant polar clusters as illustrated by the simulation snapshot in Fig.~\ref{fig:PT}.
%
Such giant polar clusters can become so big and elongated that they can can even percolate the system, as shown in panel i) of Fig.~\ref{fig:band}.
We refer to such  polar percolating structures as bands. Inside bands, rods are densely packed, point into the same direction, and exhibit  positional order. 
Notice that these bands are distinct from the bands observed in the Vicsek model, which are elongated in the direction orthogonal to the moving direction of the particles~\cite{gregoire2004}. 
More importantly, the observed polar bands are also fundamentally different from those observed in model III, where we saw that the point-like self-propelled particles 
form nematic bands, inside which 50\% of the particles move in one direction and 50\% in the opposite one. 
More importantly, our finite size study indicates that the polar patterns observed in ABR are a finite size effect that disappear for large enough systems. 
In short, several of the phases reported for $\eta \leq 0.3$  in previous ABR works~\cite{wensink2012, wensink2012b} 
such as the so-called swarming phase and the bio-turbulence phase vanish in the thermodynamical limit. 

The abrupt change in scaling of $S_1$ with $N$ in Fig.~\ref{fig:FSS} suggests that above the crossover system size $N_*$ the polar structures are no longer stable.
%
Arguably, the decay in $S_1$ with $N$ is due to the fact that rods inside polar clusters are densely packed and hold fixed positions, not being 
able to exchange neighbors in contrast to other active systems~\cite{vicsek1995,ginelli2010,marchetti2012b,abkenar2013,gregoire2004}.  
%
%
%
%
Fig.~\ref{fig:band}  shows that  $T_{agg}/T_{tot}$, {\it i.e.} the total time $T_{agg}$ the system spends in the aggregate phase with respect to the total simulation time $T_{tot}$. 
We observe that $T_{agg}$  increases with $N$, in such a way that $T_{agg}/T_{tot} \to 1$as $N\to\infty$.  This means that the probability of observing the system in an aggregate phase also increases with $N$. 
For small system sizes $N \ll N_*$ we observe moving clusters and bands. Large polar structures such as bands form, remain in the system for  quite some time, and then quickly break and reform, typically adopting a new orientation. 
As $N \to N_*$,  bands  survive for relatively short periods of time, and quickly bend and break. Interestingly, at such large system sizes other macroscopic structures start to frequently emerge.  
These new macroscopic structures -- which we refer to as {\it aggregates} --  are formed by polar clusters of rods that exert stresses on each other and exhibit vanishing polar order, see panel ii) of Fig.~\ref{fig:band}. 
In summary, for $N \sim N_*$, the system continuously transitions between highly ordered phases -- e.g. phases with either a few giant polar clusters or a band -- and aggregates, as illustrated in Fig.~\ref{fig:band}.  
For system size larger than $N_*$  bands and polar phases disappear in the thermodynamical limit, while the aggregate phase survives.  

The  transitions between aggregates and bands (or highly ordered phases) for $N \sim N_*$ 
 results from the competition between elastic energy and the impossibility of the system to sustain long-range polar order. 
For not too large system sizes, {\it i.e.} for $N\sim N_*$, the shape of the aggregates is roughly circular (Fig.~\ref{fig:band}, panel ii)) 
and  at the center of the aggregate we find one single topological defect: i.e. at the mesoscale, at the center of the aggregate we cannot define an average orientation for the rods. 
Due to the active forces,  at the center of the aggregate rods are strongly compressed, which implies that the  potentials $U_i$ adopts 
high values.  
This implies that when one of these aggregates is formed, the total elastic energy of the system $U_{tot}=\sum_{i=1}^{N}U_i$ increases. 
On the contrary, in large polar structures such as bands, rods are roughly parallel to each other and therefore are much less compressed by their neighbors, and   
the  total elastic energy is low. 
The dynamics at $N \sim N_*$ can be summarized as follows. Large polar clusters form and eventually a band emerges, 
but since the system is too big for the band to remain stable, at some point the band breaks. 
The collapse of the band gives rise to the formation of new giant polar clusters which eventually collide head on leading to a large aggregate: a process reminiscent of a traffic jam. 
The formation of the aggregate leads to a sharp increase of the total elastic energy. Let us recall that forces and torques act in such a way that they tend to minimize $U_i$. 
In short, the system relaxes by destroying the new formed aggregate, which give rise to the formation of new polar clusters and the cycle starts again. 
In larger system sizes, i.e. for  $N \gg N_*$, aggregates are more complex.

\begin{figure} 
\begin{center}
\resizebox{10cm}{!}{\rotatebox{0}{\includegraphics{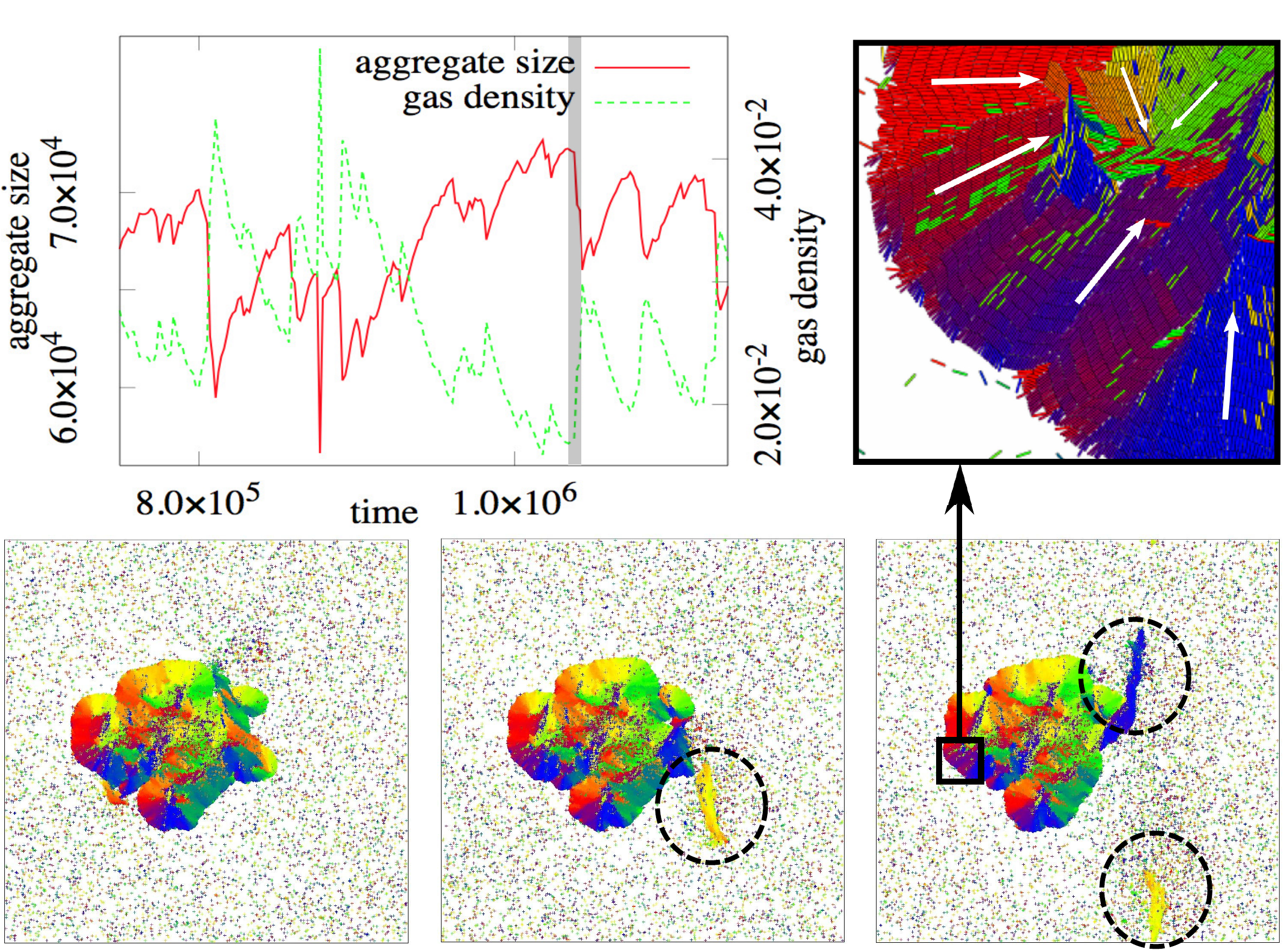}}}
\caption{[model I] 
Dynamics of an aggregate. 
Top panel: aggregate size and gas density as function of time. 
The aggregate boundary exhibits large fluctuations due to the emergence of orientational defects that lead to the detachment of large polar clusters from the aggregate. 
The three bottom panels display in chronological order one of these events. The corresponding time window is indicated by the vertical grey area in the top left panel.  Dashed circles indicate the detachment of polar clusters. The inset shows that topological defect.    
$N=80000$, $\eta=0.15$, and $\kappa=7$.
Adapted from~\cite{weitz2015}. 
%
 }
\label{fig:aggregate}
\end{center}
\end{figure}

Inside aggregates, the competition between active forces and local polar alignment leads to new physics. 
This is particularly evident for very large system sizes, i.e. for $N\gg N_*$, 
that is when aggregates are big enough to exhibit multiple topological defects of the local orientation of the rods.
Let us recall that aggregates are formed by polar clusters of rods that are trapped inside these structures.
Topological defects are areas where, at the mesoscale, as mentioned above, we cannot define an average orientation of the rods as illustrated in inset of Fig.~\ref{fig:aggregate} (areas where the arrows meet). 
In such areas, due to the active forces, rods are strongly compressed by the active push of all surrounding ABR (see inset in Fig.~\ref{fig:aggregate}). 
Since the compression is due to the presence of active forces, we refer to this phenomenon as {\it active stresses}. 
For $N\gg N_*$, we observe the emergence of multiple defects that lead to an increase of the elastic energy and the build-up of stresses. Notice that more topological defects imply larger values of the elastic energy. 
There are two clear consequence of the presence of multiple topological defects. On the one hand, aggregates are no longer roundish but rather irregular as illustrated in Fig.~\ref{fig:aggregate}. 
On the other hand, now the system can relax the elastic energy by reducing the number of topological defects. 
Notice that for $N \sim N_*$, aggregates are relatively small and exhibit one topological defect, 
and thus, the only way to eliminate the topological defects is by destroying the aggregate. 
For $N\gg N_*$, given the presence of multiple topological defects, eliminating one topological defect does not require to eliminate the aggregate.  
As a matter of fact, for very large system sizes, 
the interplay between topological defects and active stresses lead to  large fluctuations of the aggregate boundary and aggregate mass (i.e. aggregate size) as indicated in Fig.~\ref{fig:aggregate}. 
The most distinctive feature of the observed phenomenon is the large fluctuations experienced by the aggregate mass 
correspond to ejections of remarkably large macroscopic polar clusters from the aggregate, that can be as large as 10\% of the system size (i.e. involving more than $10^4$ rods), Fig.~\ref{fig:aggregate}. 
By this process, i.e. the ejection of large polar clusters, the aggregate manages to decrease its elastic energy. 
The ejected polar clusters typically dissolve while moving through the gas phase outside the aggregate, leading to a sudden increase of the gas density,  top panel in Fig.~\ref{fig:aggregate}. This results in a higher absorption rate of ABR by the aggregate that starts again to increase its mass.

\section{Discussion} \label{sec:discussion}

The comparison of models I, II, and III is insightful. 
Model II and III share the same interaction potential, which we refer here to as alignment mechanism: Eq.~(\ref{eq:evol_theta_model_II}) is identical to (\ref{eq:evol_theta_model_III}). 
Despite of this, on large scales, model II and model III display qualitatively different features. 
While model II exhibits for low noise values QLRO, as expected for an equilibrium system with continuum 
symmetry in two-dimensions, model III at low noise displays nematic LRO~\footnote{
The conclusion that nematic LRO is present in model III at low noise values is based, so far, on the numerical evidence exclusively, 
and a solid theoretical argument is still lacking. Nevertheless, it is clear that order is strongly enhanced by the active term present in model III and absence in model II, 
where particle motion is at all scales diffusive.}. 
The active term in Eq.~(\ref{eq:evol_theta_model_III}), $v_0 \mathbf{V}(\theta_i)$, 
makes possible the emergence of stable nematic bands and an homogeneous nematically ordered phase (phase I). 
In summary, in model II we observe local nematic order and QLRO at large scales, while 
in model III we find again local nematic order and LRO at large scales. 
Importantly, at the meso and macroscopic sale model II and III do not exhibit polar order. 

In model I, on the other hand,  nematic order is never observed (at least up to packing fractions $0.3$). 
The difference between model I and model III is due to the presence of the repulsive term $-\nabla U_i$ in the equation for $\dot{\mathbf{x}}_i$. 
We have learned that the  addition of  volume exclusion interactions in the evolution of the position of the particle has a dramatic effect on the large-scale properties 
of the system. 
At small system sizes and for large enough aspect ratios (or equivalently packing fractions) particles self-organize 
into large moving polar clusters and the system displays global polar order. Moreover, the observed bands are polar instead of being nematic as the ones observed in model III. 
However, the performed finite size study reveals that polar order (or any orientational order) vanishes in the thermodynamical limit (at least up to packing fractions $0.3$). 
Thus, 
from model I we learn that volume exclusion effects induce local polar and absence of orientational order at large scale.   
Despite the lack of orientation order, model I undergoes a genuine phase-separation. 
The phase-separation process, as well as the phase-separated phase are qualitatively different from the ones observed in model III, where 
the system spontaneous phase separates in the form of nematically ordered, high-density bands (referred above as phase II and III).  
In model I, phase separation starts with the formation if giant polar clusters~\cite{peruani2013} that jam to form what we have called aggregates. 
The most distinctive feature of these aggregates is the constant ejection from the aggregate of thousands of particles in the form of densely packed and polarly ordered clusters, 
which leads to large fluctuation of the aggregate size and its boundary. 
This occurs due to the combined effect of an effective alignment mechanism (also present in model III) and the active pushing (or stresses) acting among the particles 
that requires not only an active force but also a term $-\nabla U_i$ in the evolution of the position of the particles. 

The combination of these elements (active stresses and alignment) is not present in other active system, making the physics of active Brownian rods unique. 
For instance, self-propelled disks (SPD) exhibit active stresses but no alignment among the SPD~\cite{fily2012,buttinoni2013,redner2013}. 
Thus, phase separation in SPD resembles a classical coarsening process~\cite{fily2012,buttinoni2013,redner2013} which can be described by an effective Cahn-Hilliard equation~\cite{speck2014}.    
Finally, it important to stress that the phase separation in model I is different from the one observed in active system with an alignment mechanism and a density-dependent speed~\cite{marchetti2012b,peruani2011}. Here, while alignment among particles is present, there are no active stresses~\footnote{The slow down of the particles due to a density-dependent speed, does not imply active pushing.}.   

In summary, we have characterized the large-scale properties of ABR  for  packing fractions smaller than or equal to $0.3$.   
We have shown that large-scale (orientational) order patterns cannot exist in the thermodynamic limit for (physical) ABR ({\it i.e.} in model I) 
due to  combined effect of active stresses and alignment, which  leads to exciting new physics. 
In particular, we have seen that 
the interplay of these two elements (active stresses and alignment) gives rise to a  novel highly fluctuating phase-separated phase characterized by the ejection of giant polar clusters. 
We point out that  the presented evidence  cannot be used to preclude the emergence of orientational order 
at higher packing fractions. 
Let us recall that even for Brownian rods we expect for large enough packing fractions some kind of isotropic-nematic transition. 
In simulations with ABR at high packing a transition to laning has been reported~\cite{mccandlish2012,wensink2012b,abkenar2013,kuan2014}. 
The observed laning phase may remain even in the thermodynamic limit, though its existence should be confirmed by a careful finite size study. 
Finally, the model presented in~\cite{abkenar2013} for penetrable ABR suggested a possible crossover between the large-scale properties exhibited 
by model I and III. While this scenario cannot be {\it a priori} excluded in~\cite{abkenar2013},  a finite size study may reveal some of the reported phases vanish in the thermodynamical limit.




%
%
%
%




\bibliographystyle{apsrev}


\begin{thebibliography}{99}

\expandafter\ifx\csname natexlab\endcsname\relax\def\natexlab#1{#1}\fi
\expandafter\ifx\csname bibnamefont\endcsname\relax
  \def\bibnamefont#1{#1}\fi
\expandafter\ifx\csname bibfnamefont\endcsname\relax
  \def\bibfnamefont#1{#1}\fi
\expandafter\ifx\csname citenamefont\endcsname\relax
  \def\citenamefont#1{#1}\fi
\expandafter\ifx\csname url\endcsname\relax
  \def\url#1{\texttt{#1}}\fi
\expandafter\ifx\csname urlprefix\endcsname\relax\def\urlprefix{URL }\fi
\providecommand{\bibinfo}[2]{#2}
\providecommand{\eprint}[2][]{\url{#2}}

\bibitem{peruani2006}
\bibinfo{author}{\bibfnamefont{F.}~\bibnamefont{Peruani}},
  \bibinfo{author}{\bibfnamefont{A.}~\bibnamefont{Deutsch}}, \bibnamefont{and}
  \bibinfo{author}{\bibfnamefont{M.}~\bibnamefont{B{\"a}r}},
  \bibinfo{journal}{Phys. Rev. E} \textbf{\bibinfo{volume}{74}},
  \bibinfo{pages}{030904(R)} (\bibinfo{year}{2006}).

\bibitem{yang2010}
\bibinfo{author}{\bibfnamefont{Y.}~\bibnamefont{Yang}},
  \bibinfo{author}{\bibfnamefont{V.}~\bibnamefont{Marceau}}, \bibnamefont{and}
  \bibinfo{author}{\bibfnamefont{G.}~\bibnamefont{Gompper}},
  \bibinfo{journal}{Phys. Rev. E} \textbf{\bibinfo{volume}{82}},
  \bibinfo{pages}{031904} (\bibinfo{year}{2010}).

\bibitem{baskaran2008}
\bibinfo{author}{\bibfnamefont{A.}~\bibnamefont{Baskaran}} \bibnamefont{and}
  \bibinfo{author}{\bibfnamefont{M.}~\bibnamefont{Marchetti}},
  \bibinfo{journal}{Phys. Rev. Lett.} \textbf{\bibinfo{volume}{101}},
  \bibinfo{pages}{268101} (\bibinfo{year}{2008}).

\bibitem{peruani2012}
\bibinfo{author}{\bibfnamefont{F.}~\bibnamefont{Peruani}},
  \bibinfo{author}{\bibfnamefont{J.}~\bibnamefont{Starruss}},
  \bibinfo{author}{\bibfnamefont{V.}~\bibnamefont{Jakovljevic}},
  \bibinfo{author}{\bibfnamefont{L.}~\bibnamefont{Sogaard-Andersen}},
  \bibinfo{author}{\bibfnamefont{A.}~\bibnamefont{Deutsch}}, \bibnamefont{and}
  \bibinfo{author}{\bibfnamefont{M.}~\bibnamefont{B{\"a}r}},
  \bibinfo{journal}{Phys. Rev. Lett.} \textbf{\bibinfo{volume}{108}},
  \bibinfo{pages}{098102} (\bibinfo{year}{2012}).
  
  
  
\bibitem{starruss2012}
\bibinfo{author}{\bibfnamefont{J.}~\bibnamefont{Starruss}},
  \bibinfo{author}{\bibfnamefont{F.}~\bibnamefont{Peruani}},
  \bibinfo{author}{\bibfnamefont{V.}~\bibnamefont{Jakovljevic}},
  \bibinfo{author}{\bibfnamefont{L.}~\bibnamefont{Sogaard-Andersen}},
  \bibinfo{author}{\bibfnamefont{A.}~\bibnamefont{Deutsch}}, \bibnamefont{and}
  \bibinfo{author}{\bibfnamefont{M.}~\bibnamefont{B{\"a}r}},
  \bibinfo{journal}{Interface focus} \textbf{\bibinfo{volume}{2}},
  \bibinfo{pages}{774} (\bibinfo{year}{2012}).

\bibitem{kudrolli2008}
\bibinfo{author}{\bibfnamefont{A.}~\bibnamefont{Kudrolli}},
  \bibinfo{author}{\bibfnamefont{G.}~\bibnamefont{Lumay}},
  \bibinfo{author}{\bibfnamefont{D.}~\bibnamefont{Volfson}}, \bibnamefont{and}
  \bibinfo{author}{\bibfnamefont{L.}~\bibnamefont{Tsimring}},
  \bibinfo{journal}{Phys. Rev. Lett.} \textbf{\bibinfo{volume}{100}},
  \bibinfo{pages}{058001} (\bibinfo{year}{2008}).

\bibitem{kudrolli2010}
\bibinfo{author}{\bibfnamefont{A.}~\bibnamefont{Kudrolli}},
  \bibinfo{journal}{Phys. Rev. Lett.} \textbf{\bibinfo{volume}{104}},
  \bibinfo{pages}{088001} (\bibinfo{year}{2010}).

\bibitem{paxton2004}
\bibinfo{author}{\bibfnamefont{W.}~\bibnamefont{Paxton}} \bibnamefont{and}
  \bibinfo{author}{\bibnamefont{al.}}, \bibinfo{journal}{J. Am. Chem. Soc.}
  \textbf{\bibinfo{volume}{126}}, \bibinfo{pages}{13424}
  (\bibinfo{year}{2004}).

\bibitem{mano2005}
\bibinfo{author}{\bibfnamefont{N.}~\bibnamefont{Mano}} \bibnamefont{and}
  \bibinfo{author}{\bibfnamefont{A.}~\bibnamefont{Heller}},
  \bibinfo{journal}{J. Am. Chem. Soc.} \textbf{\bibinfo{volume}{127}},
  \bibinfo{pages}{11574} (\bibinfo{year}{2005}).

  \bibitem{wensink2012}
\bibinfo{author}{\bibfnamefont{H.~H.} \bibnamefont{Wensink}},
  \bibinfo{author}{\bibfnamefont{J.}~\bibnamefont{Dunkel}},
  \bibinfo{author}{\bibfnamefont{S.}~\bibnamefont{Heidenreich}},
  \bibinfo{author}{\bibfnamefont{K.}~\bibnamefont{Drescher}},
  \bibinfo{author}{\bibfnamefont{R.~E.} \bibnamefont{Goldstein}},
  \bibinfo{author}{\bibfnamefont{H.}~\bibnamefont{L{\"o}wen}},
  \bibnamefont{and} \bibinfo{author}{\bibfnamefont{J.~M.}
  \bibnamefont{Yeomans}}, \bibinfo{journal}{Proc. Natl. Acad. Sci. USA}
  \textbf{\bibinfo{volume}{109}}, \bibinfo{pages}{14308}
  (\bibinfo{year}{2012}).

\bibitem{dunkel2013} J. Dunkel, S. Heidenreich, K. Drescher, H.H. Wensink, M. B{\"a}r, and R.E. Goldstein, Phys. Rev. Lett. {\bf 110}, 228102 (2013). 
  
\bibitem{zhang2010} H.P. Zhang, A. Be'er, E.-L. Florin, and H.L. Swinney, Proc. Natl. Acad. Sci. USA 107, 13626-13630 (2010)


\bibitem{schaller2010}
\bibinfo{author}{\bibfnamefont{V.}~\bibnamefont{Schaller}},
  \bibinfo{author}{\bibfnamefont{C.}~\bibnamefont{Weber}},
  \bibinfo{author}{\bibfnamefont{C.}~\bibnamefont{Semmrich}},
  \bibinfo{author}{\bibfnamefont{E.}~\bibnamefont{Frey}}, \bibnamefont{and}
  \bibinfo{author}{\bibfnamefont{A.}~\bibnamefont{Bausch}},
  \bibinfo{journal}{Nature} \textbf{\bibinfo{volume}{467}}, \bibinfo{pages}{73}
  (\bibinfo{year}{2010}).

\bibitem{sumino2012}
\bibinfo{author}{\bibfnamefont{Y.}~\bibnamefont{Sumino}},
  \bibinfo{author}{\bibfnamefont{K.}~\bibnamefont{Nagai}},
  \bibinfo{author}{\bibfnamefont{Y.}~\bibnamefont{Shitaka}},
  \bibinfo{author}{\bibfnamefont{D.}~\bibnamefont{Tanaka}},
  \bibinfo{author}{\bibfnamefont{K.}~\bibnamefont{Yoshikawa}},
  \bibinfo{author}{\bibfnamefont{H.}~\bibnamefont{Chat{\'e}}},
  \bibnamefont{and} \bibinfo{author}{\bibfnamefont{K.}~\bibnamefont{Oiwa}},
  \bibinfo{journal}{Nature} \textbf{\bibinfo{volume}{483}},
  \bibinfo{pages}{448} (\bibinfo{year}{2012}).

\bibitem{peruani2011}
\bibinfo{author}{\bibfnamefont{F.}~\bibnamefont{Peruani}},
  \bibinfo{author}{\bibfnamefont{T.}~\bibnamefont{Klauss}},
  \bibinfo{author}{\bibfnamefont{A.}~\bibnamefont{Deutsch}}, \bibnamefont{and}
  \bibinfo{author}{\bibfnamefont{A.}~\bibnamefont{Voss-Boehme}},
  \bibinfo{journal}{Phys. Rev. Lett.} \textbf{\bibinfo{volume}{106}},
  \bibinfo{pages}{128101} (\bibinfo{year}{2011}).

  \bibitem{marchetti2012b}
\bibinfo{author}{\bibfnamefont{F.~D.~C.} \bibnamefont{Farrell}},
  \bibinfo{author}{\bibfnamefont{M.~C.} \bibnamefont{Marchetti}},
  \bibinfo{author}{\bibfnamefont{D.}~\bibnamefont{Marenduzzo}},
  \bibnamefont{and} \bibinfo{author}{\bibfnamefont{J.}~\bibnamefont{Tailleur}},
  \bibinfo{journal}{Phys. Rev. Lett.} \textbf{\bibinfo{volume}{108}},
  \bibinfo{pages}{248101} (\bibinfo{year}{2012}).
  
\bibitem{grossmann2012} R. Gro{\ss}mann, L. Schimansky-Geier, and P. Romanczuk, New J. Phys. {\bf 14}, 073033 (2012).

\bibitem{weitz2015} S. Weitz, A. Deutsch, and F. Peruani
Phys. Rev. E 92, 012322 (2015).


\bibitem{onsager1949}
\bibinfo{author}{\bibfnamefont{L.}~\bibnamefont{Onsager}},
  \bibinfo{journal}{Ann. N.Y. Acad. Sci.} \textbf{\bibinfo{volume}{51}},
  \bibinfo{pages}{627} (\bibinfo{year}{1949}).



\bibitem{vicsek2012}
\bibinfo{author}{\bibfnamefont{T.}~\bibnamefont{Vicsek}} \bibnamefont{and}
  \bibinfo{author}{\bibfnamefont{A.}~\bibnamefont{Zafeiris}},
  \bibinfo{journal}{Physics Reports} \textbf{\bibinfo{volume}{517}},
  \bibinfo{pages}{71} (\bibinfo{year}{2012}).

\bibitem{marchetti2013} 
\bibinfo{author}{\bibfnamefont{M.~C.} \bibnamefont{Marchetti}},
  \bibinfo{author}{\bibfnamefont{J.~F.} \bibnamefont{Joanny}},
  \bibinfo{author}{\bibfnamefont{S.}~\bibnamefont{Ramaswamy}},
  \bibinfo{author}{\bibfnamefont{T.~B.} \bibnamefont{Liverpool}},
  \bibinfo{author}{\bibfnamefont{M.~R.} \bibnamefont{J.~Prost}},
  \bibnamefont{and} \bibinfo{author}{\bibfnamefont{R.~A.} \bibnamefont{Simha}},
  \bibinfo{journal}{Rev. Mod. Phys.} \textbf{\bibinfo{volume}{85}},
  \bibinfo{pages}{1143} (\bibinfo{year}{2013}).

\bibitem{ginelli2010}
\bibinfo{author}{\bibfnamefont{F.}~\bibnamefont{Ginelli}},
  \bibinfo{author}{\bibfnamefont{F.}~\bibnamefont{Peruani}},
  \bibinfo{author}{\bibfnamefont{M.}~\bibnamefont{B{\"a}r}}, \bibnamefont{and}
  \bibinfo{author}{\bibfnamefont{H.}~\bibnamefont{Chat{\'e}}},
  \bibinfo{journal}{Phys. Rev. Lett.} \textbf{\bibinfo{volume}{104}},
  \bibinfo{pages}{184502} (\bibinfo{year}{2010}).


\bibitem{wensink2012b}
\bibinfo{author}{\bibfnamefont{H.~H.} \bibnamefont{Wensink}} \bibnamefont{and}
  \bibinfo{author}{\bibfnamefont{H.}~\bibnamefont{L{\"o}wen}},
  \bibinfo{journal}{J. Phys.: Condens. Matt.} \textbf{\bibinfo{volume}{24}},
  \bibinfo{pages}{464130} (\bibinfo{year}{2012}).

  \bibitem{abkenar2013}
\bibinfo{author}{\bibfnamefont{M.}~\bibnamefont{Abkenar}},
  \bibinfo{author}{\bibfnamefont{K.}~\bibnamefont{Marx}},
  \bibinfo{author}{\bibfnamefont{T.}~\bibnamefont{Auth}}, \bibnamefont{and}
  \bibinfo{author}{\bibfnamefont{G.}~\bibnamefont{Gompper}},
  \bibinfo{journal}{Phys. Rev. E} \textbf{\bibinfo{volume}{88}},
  \bibinfo{pages}{062314} (\bibinfo{year}{2013}).

  \bibitem{mccandlish2012}
\bibinfo{author}{\bibfnamefont{S.~R.} \bibnamefont{McCandlish}},
  \bibinfo{author}{\bibfnamefont{A.}~\bibnamefont{Baskaran}}, \bibnamefont{and}
  \bibinfo{author}{\bibfnamefont{M.~F.} \bibnamefont{Hagan}},
  \bibinfo{journal}{Soft Matter} \textbf{\bibinfo{volume}{8}},
  \bibinfo{pages}{2527} (\bibinfo{year}{2012}).



%
%
%

%
%
%



 \bibitem{peruani2008}
\bibinfo{author}{\bibfnamefont{F.}~\bibnamefont{Peruani}},
  \bibinfo{author}{\bibfnamefont{A.}~\bibnamefont{Deutsch}}, \bibnamefont{and}
  \bibinfo{author}{\bibfnamefont{M.}~\bibnamefont{B{\"a}r}},
  \bibinfo{journal}{Eur. Phys. J. Special Topics}
  \textbf{\bibinfo{volume}{157}}, \bibinfo{pages}{111} (\bibinfo{year}{2008}).

 \bibitem{vicsek1995}
\bibinfo{author}{\bibfnamefont{T.}~\bibnamefont{Vicsek}},
  \bibinfo{author}{\bibfnamefont{E.}~\bibnamefont{A.~Czirok}},
  \bibinfo{author}{\bibfnamefont{E.~B.} \bibnamefont{Jacob}},
  \bibinfo{author}{\bibfnamefont{I.}~\bibnamefont{Cohen}}, \bibnamefont{and}
  \bibinfo{author}{\bibfnamefont{O.}~\bibnamefont{Shochet}},
  \bibinfo{journal}{Phys. Rev. Lett.} \textbf{\bibinfo{volume}{75}},
  \bibinfo{pages}{1226} (\bibinfo{year}{1995}).



  
  \bibitem{berezinskii_destruction_1971}
\bibinfo{author}{\bibfnamefont{V.}~\bibnamefont{Berezinskii}},
  \bibinfo{journal}{Sov. Phys. JETP} \textbf{\bibinfo{volume}{32}},
  \bibinfo{pages}{493} (\bibinfo{year}{1971}).

  \bibitem{kosterlitz_ordering_1973}
\bibinfo{author}{\bibfnamefont{J.~M.} \bibnamefont{Kosterlitz}}
  \bibnamefont{and} \bibinfo{author}{\bibfnamefont{D.~J.}
  \bibnamefont{Thouless}}, \bibinfo{journal}{J. Phys. C (Solid State)}
  \textbf{\bibinfo{volume}{6}}, \bibinfo{pages}{1181} (\bibinfo{year}{1973}).

\bibitem{mermin1966}
\bibinfo{author}{\bibfnamefont{N.~D.} \bibnamefont{Mermin}} \bibnamefont{and}
  \bibinfo{author}{\bibfnamefont{H.}~\bibnamefont{Wagner}},
  \bibinfo{journal}{Phys. Rev. Lett.} \textbf{\bibinfo{volume}{17}},
  \bibinfo{pages}{1133} (\bibinfo{year}{1966}).

  
\bibitem{delasheras2013} D. de las Heras, Y. Martinez-Raton, L. Mederos, E. Velasco, J. Molecular Liquids {\bf 185}, 13-19 (2013).
  
\bibitem{doi1986} M. Doi, and S.F. Edwards, \emph{The theory of polymer dynamics.} (Oxford Univ. Press, New York, 1986).

\bibitem{levine2004} A.J. Levine, T. Liverpool, and F. MacKintosch, Phys. Rev. E {\bf 69}, 021503 (2004).



 

 

\bibitem{lebwohl1973} P.A. Lebwohl, G. Lasher, Phys. Rev. A {\bf 6}, 426 (1973). 

\bibitem{farinas2010} A.I. Fari{n}as-S{\'a}nchez et al., Cond. Matter Phys. {\bf 13}, 13601 (2010). 

\bibitem{robert} R. Grossmann, F. Peruani, M. B{\"ar}, arXiv.1504.01694 (2015).

\bibitem{peruani2013}
\bibinfo{author}{\bibfnamefont{F.}~\bibnamefont{Peruani}} \bibnamefont{and}
  \bibinfo{author}{\bibfnamefont{M.}~\bibnamefont{B{\"a}r}},
  \bibinfo{journal}{New J. Phys.} \textbf{\bibinfo{volume}{15}},
  \bibinfo{pages}{065009} (\bibinfo{year}{2013}).
  
  \bibitem{peruani2010} F. Peruani, L. Schimansky-Geier, M. B{\"a}r, Eur. Phys. J. Special Topics {\bf 191}, 173-185 (2010). 

\bibitem{gregoire2004}
\bibinfo{author}{\bibfnamefont{G.}~\bibnamefont{Gr{\'e}goire}}
  \bibnamefont{and}
  \bibinfo{author}{\bibfnamefont{H.}~\bibnamefont{Chat{\'e}}},
  \bibinfo{journal}{Phys. Rev. Lett.} \textbf{\bibinfo{volume}{92}},
  \bibinfo{pages}{025702} (\bibinfo{year}{2004}).

 
 \bibitem{chate2006} H. Chat{\'e}, F. Ginelli, and R. Montagne, Phys. Rev. Lett. 96, 180602 (2006).
 
\bibitem{buttinoni2013} I. Buttinoni, J. Bialk{\'e}, F. K{\"u}mmel, H. L{\"o}wen, C. Bechinger, T. Speck, Phys. Rev. Lett. {\bf 110}, 238301 (2013). 


\bibitem{fily2012}
\bibinfo{author}{\bibfnamefont{Y.}~\bibnamefont{Fily}} \bibnamefont{and}
  \bibinfo{author}{\bibfnamefont{M.}~\bibnamefont{Marchetti}},
  \bibinfo{journal}{Phys. Rev. Lett.} \textbf{\bibinfo{volume}{108}},
  \bibinfo{pages}{235702} (\bibinfo{year}{2012}).

  \bibitem{redner2013}
\bibinfo{author}{\bibfnamefont{G.}~\bibnamefont{Redner}},
  \bibinfo{author}{\bibfnamefont{M.}~\bibnamefont{Hagan}}, \bibnamefont{and}
  \bibinfo{author}{\bibfnamefont{A.}~\bibnamefont{Baskaran}},
  \bibinfo{journal}{Phys. Rev. Lett.} \textbf{\bibinfo{volume}{110}},
  \bibinfo{pages}{055701} (\bibinfo{year}{2013}).


\bibitem{speck2014} T. Speck, J. Bialk{\'e}, A.M. Menzel, and H. L{\"o}wen, Phys. Rev. Lett. {\bf 110}, 218304 (2014). 

\bibitem{kuan2014} H.-S. Kuan, R. Blackwell, M.A. Glaser, M.D. Betterton,  arXiv:1407.4842 (2014).






       
\end{thebibliography}

\end{document}